\documentclass[twocolumn,shortnote]{jpsj2} 
\usepackage{bm}
\usepackage{color}

\title{The London Equation in Quantum Spin Hall System with Electronic Correlation}

\author{Jun \textsc{Goryo}
\thanks{E-mail address: jungoryo@iis.u-tokyo.ac.jp} and 
Nobuki \textsc{Maeda}$^{1}$
\thanks{ E-mail address: maeda@particle.sci.hokudai.ac.jp}}

\inst{Institute of Industrial Science, the University of Tokyo, 4-6-1, Komaba, Meguro, Tokyo, Japan \\
$^{1}$ Department of Physics, Hokkaido University, Sapporo, Japan}

\kword{quantum spin Hall effect, Kane-Mele model, electron correlation, effective field theory, London equation, BF term}

\begin{document}
\maketitle


The Kane-Mele (KM) model is proposed to describe the quantum spin Hall effect (QSHE) of 
electrons on the two-dimensional honeycomb lattice\cite{KM1,KM2}. 
Here, we will show that, in a certain parameter region, the London equation is obtained 
from the effective field theory of the KM model with an electronic correlation\cite{Shitade,Pesin-Balents,Rachel-LeHur}. 
We use $\hbar=c=1$ unit and the Minkovskian metric $g^{\mu\nu}=diag(1,-1,-1)$, where 
$\mu,\nu=0,x,y$.

We consider a layered honeycomb lattice system in which each layer is described by the KM model. We assume that 
interlayer coupling is negligibly small.
One of the essential ingredients of the KM model\cite{KM1,KM2} is intrinsic spin-orbit (SO) coupling 
$\lambda_{\rm SO}$, which generates an effective magnetic field depending on spin as well as an excitation gap $\Delta=3\sqrt{3} \lambda_{\rm SO}$ 
to the electron\cite{KM2}. Thus, the term enables quantization of the spin Hall conductivity (SHC)\cite{KM1,KM2}, 
\begin{eqnarray}
\sigma_{xy}^s=\frac{e}{2\pi d}\frac{\Delta}{|\Delta|}, 
\label{SHC}
\end{eqnarray}
where $d$ is the interlayer distance. The model can also have the Rashba extrinsic SO coupling $\lambda_R$, which 
breaks the inversion symmetry and is induced by an electric field perpendicular to the honeycomb lattice  plane. The term also breaks the conservation of electron spin $s_z/2$. Hereafter, we consider the case of $\lambda_R=0$. 

We add the on-site Coulomb repulsion $U>0$. The Hamiltonian per a layer is
\begin{eqnarray}
H&=&t\sum_{<ij>}c^\dagger_i c_j + i \lambda_{\rm SO} \sum_{<<ij>>}\nu_{ij} c^{\dagger}_i s_z c_j 
\nonumber\\
&&+ U\sum_i n_{i \uparrow}n_{i \downarrow},  
\label{latticeH}
\end{eqnarray}
where $c_i$ ($c^\dagger_i$) is the annihilation (creation) operator of an electron with spin at the $i$-th site and $t$ is nearest neighbor hopping. The second term is the intrinsic SO term consisting of next nearest neighbor hopping, and 
$\nu_{ij}=\frac{2}{\sqrt{3}} (\hat{\bm d}_1\times \hat{\bm d}_2)_z=\pm1$, where  $\hat{\bm d}_1$ and  $\hat{\bm d}_2$ are unit vectors along the two bonds 
where the electron moving from site $j$ to $i$ passes.

Let us discuss how to deal with the electron correlation $U$. 
On-site coulomb repulsion can be written by the on-site spin-spin interaction: 
\begin{eqnarray}
U n_{i\uparrow}n_{i\downarrow}=\frac{U}{2}(n_{i\uparrow}+n_{i\downarrow}) - \frac{U}{6}(c^\dagger_i \vec{s} c_i)^2. 
\end{eqnarray}
The first term merely gives the renormalization for the chemical potential and can be neglected. 
We introduce the auxiliary field $\vec{\varphi}_i$, which is a three-component vector in the spin space, 
and use the Stratonovich-Hubbard transformation,\cite{Mahan} 
$H \rightarrow H_{SH}=H+\Delta H$, where 
\begin{eqnarray}
\Delta H&=&\frac{U}{6}\sum_i (c^\dagger_i \vec{s} c_i - \frac{3}{2U} \vec{\varphi}_i)^2, 
\label{SHgauss}
\\
H_{SH}&=&t\sum_{<ij>}c^\dagger_i c_j + i \lambda_{\rm SO} \sum_{<<ij>>}\nu_{ij} c^{\dagger}_i s_z c_j 
\nonumber\\
&&-\sum_i \vec{\varphi}_i \cdot c_i^\dagger \frac{\vec{s}}{2} c_i + \frac{3}{8U}\sum_i |\vec{\varphi}_i|^2. 
\label{latticeHSH}
\end{eqnarray}
The spin-spin interaction is eliminated in appearance, but we have coupling between $\vec{\varphi}_i$ and  the electron spin, and a quadratic term of $\vec{\varphi}_i$ instead.

We consider the continuum limit and take into account the low-energy electronic 
excitations around $K$ and $K^\prime$ points in the Brillouin Zone \cite{KM1,KM2}. 
We introduce the electromagnetic $U(1)$ gauge field 
$A_\mu$ and $SU(2)$ spin gauge field $\vec{a}_\mu$ via the 
covariant derivative, 
\begin{eqnarray}
i D_\mu= i \partial_\mu - e A_\mu + \vec{a}_\mu \cdot \frac{\vec{s}}{2},  
\end{eqnarray}
where $\vec{a}_0=\vec{\varphi}$ (the auxiliary field in the continuum limit) and $\vec{\bm a}$ is a constant external field introduced artificially to estimate the spin current. 
We define a parameter 
\begin{eqnarray}
g=\frac{4 U a^2d}{3}, 
\label{g}
\end{eqnarray}
where $a$ is the lattice constant, and the microscopic Lagrangian density is
\begin{eqnarray}
{\cal{L}}&=&\Psi^\dagger \left\{i D_0 - iv (D_x \tau_z \sigma_x + D_y \sigma_y) 
+ \Delta \tau_z \sigma_z s_z \right\}\Psi
\nonumber\\
&&
+ \frac{\epsilon_0 E^2}{2}- \frac{B^2}{2\mu_0} -\frac{1}{2g}|\vec{a_0}|^{2},   
\label{microL}
\end{eqnarray}
where $\Psi=\Psi_{\tau\sigma s}$ is the eight-component Fermion field labeled by the eigenvalues of the diagonal components of valley spin $\vec{\tau}$, sublattice spin $\vec{\sigma}$ and real spin $\vec{s}/2$. The parameter $v$ is the Fermi velocity when the system is in the metallic state, and $\epsilon_0$ and $\mu_0$ denote the dielectric constant and magnetic permeability, respectively.  
Note that, except for the last term, the Lagrangian (\ref{microL}) possesses the $U(1)_{\rm em} \times U(1)_z$ local gauge symmetry.  The $SU(2)$ gauge symmetry is broken down to $U(1)_z$, since the SO term contains $s_z$. 

The calculation shown below is similar to that presented in Ref. 7, although the physical meaning of the spin gauge field is different. We integrate out $\Psi$ and obtain the one-loop effective Lagrangian for the gauge fields in the low-energy (long-wavelength) region compared with $\Delta$ ($\Delta^{-1}$).
The result is\cite{GMI} 
\begin{eqnarray}
{\cal{L}}_{\rm eff}&=&-\frac{1}{2g}a_0^{z2}+{\cal{L}}_{\rm ind}, 
\label{Leff}
\\
{\cal{L}}_{\rm ind}&=&\sigma_{xy}^s \epsilon^{\mu\rho\nu} a_\mu^z \partial_\rho A_\nu 
+ \frac{\epsilon E^2}{2} - \frac{B^2}{2\mu}+\frac{\delta \epsilon}{8e^2}({\bm \nabla} a_0^z)^2 
\nonumber\\
&&+ ({\rm terms~independent~of}~a_\mu^z~{\rm and}~A_\mu),  
\label{Lind}
\end{eqnarray}
where ${\cal{L}}_{\rm ind}$ stands for the induced part of the effective Lagrangian. The first term in Eq. (\ref{Lind}) is the BF term\cite{BF}. The coefficient is the quantized SHC given in eq. (\ref{SHC}). Note that only $a_\mu^z$ couples to the electromagnetic gauge fields, because the $SU$(2) symmetry 
is broken down to the $U(1)_z$ symmetry by SO coupling. The Maxwell term is renormalized 
as\cite{Semenoff-Sodano-Wu,GMI}
\begin{eqnarray}
\epsilon&=&\epsilon_0+\delta \epsilon
\nonumber\\
&=&\epsilon_0 + \frac{e^2}{6 \pi |\Delta| d}, 
\label{epsilon}\\
\frac{1}{\mu}&=&\frac{1}{\mu_0} + \frac{e^2 v^2}{6 \pi |\Delta| d}. 
\label{mu}
\end{eqnarray}
By using the parameters in Table \ref{table} and the relations $\frac{e^2}{4\pi\epsilon_0}\simeq 1/137$ and $\epsilon_0 \mu_0=1$, we obtain 
$\epsilon_0/\frac{e^2}{6 \pi |\Delta|d}=0.5$ and $\mu_0 \cdot \frac{e^2v^2}{6 \pi |\Delta|d}=2 \times 10^{-8}$, i.e., $\mu\simeq\mu_0$.
The elastic term for  $a_0^z(=\varphi^z)$ is also induced. 
We can recognize that any potential terms 
(i.e., zeroth-order terms with respect to the derivative $\partial_\mu$) of $A_\mu$ and also $a_\mu^z$ in ${\cal{L}}_{\rm ind}$ 
are absent because of the presence of $U(1)_{\rm em} \times U(1)_z$ gauge symmetry in the Fermionic part of the microscopic Lagrangian (\ref{microL}). 
Thus, the low-energy and long-wavelength physics of $A_\mu$ and $a_\mu^z$ 
is described definitely by eq. (\ref{Leff}). 

We consider the static magnetic response. The equations of motion obtained from eq. (\ref{Leff}) are
\begin{eqnarray}
\frac{\delta \epsilon}{4e^2} \nabla^2 a_0^z+\frac{1}{g}a_0^z&=&\sigma_{xy}^s B, 
\label{poisson+1/g}\\
\frac{1}{\mu} \sum_{j=x,y} \epsilon_{ij} \nabla_j B &=&\sigma_{xy}^s \sum_{j=x,y} \epsilon_{ij} \nabla_j a_0^z.   
\label{Maxwell}
\end{eqnarray}
If we obtain $B$ from eqs. (\ref{poisson+1/g}) and (\ref{Maxwell}), we can find that the term $a_0^z/g$ in eq. (\ref{poisson+1/g}) can be neglected when 
\begin{equation}
\sigma_{xy}^{s2}\gg\frac{1}{\mu g} (>0). 
\label{condition}
\end{equation}
Going back to the microscopic Lagrangian (\ref{microL}), we may disregard the quadratic term in this case, 
and thus, $a_0^z$ can be regarded as ``the spin chemical potential". 

Here, we make {\it two crucial assumptions}: (i) eq. (\ref{condition}) is satisfied 
and (ii) the system is not in the topological Mott insulating phase\cite{Pesin-Balents} but in 
the topological band insulating phase, which exhibits the quantum spin Hall effect and its low-energy and 
long-wavelength physics is described by the effective Lagrangian eq. (\ref{Leff}) when the spin is conserved (e.g., $\lambda_R=0$). 
To realize (ii), $U$ should not be too large.\cite{Pesin-Balents,Rachel-LeHur} Thus, a large $a$ is favorable to realize (i) and (ii) simultaneously 
[see eq. (\ref{g})]. Note that a large $d$ is unfavorable since $\sigma_{xy}^s$  is suppressed [see eq. (\ref{SHC})].  

As mentioned, the $1/g$ term in eq. (\ref{poisson+1/g}) can be neglected under condition eq. (\ref{condition}).
The {\it r.h.s.} of eq. (\ref{poisson+1/g}) is a result of the fact that the spin density is proportional to the magnetic field $B$ owing to the BF term. 
It resembles the Zeeman effect, but the essential difference is that the coefficient is not the Bohr magneton but the SHC. 
The {\it r.h.s} of eq. (\ref{Maxwell}) is a result of the fact that, owing to the BF term, the electric current flows perpendicular 
to the gradient of the spin chemical potential $a_0^z$. This may be called the dual quantized spin Hall effect (dual QSHE). 
Taking the rotation of both sides of eq. (\ref{Maxwell}) and using eq. (\ref{poisson+1/g}) (note, again, that we are neglecting the $1/g$ term), we obtain the London equation\cite{GMI,Diamantini},
\begin{eqnarray}
\frac{1}{\mu}\nabla^2 B =\frac{4e^2 \sigma_{xy}^{s2}}{\delta \epsilon} B. 
\label{London}
\end{eqnarray}
Let us check condition (\ref{condition}). 
By using the parameters for the topological band insulator Na$_2$IrO$_3$\cite{Shitade} in Table \ref{table}, we obtain 
$\sigma_{xy}^{s2}\mu g  \simeq 7.0 \times10^{-6}$, i.e., eq. (\ref{condition}) is not satisfied. 
We note that, if we have a superlattice structure with $a=10^4\AA$, eq. (\ref{condition}) is 
satisfied since $\sigma_{xy}^{s2}\mu g\simeq 7$. 
   

The physical implication of eq. (\ref{London}) is the Meissner effect. 
The penetration depth of the magnetic field is estimated using the parameters in Table \ref{table} as
$
\lambda_{pen.}=2 \pi (2 e \sigma_{xy}^{s})^{-1}\delta \epsilon^{1/2}\mu^{-1/2} \simeq 3000~\AA,
$
which is as short as the typical value of the order of 1000 $\AA$ for superconductors. 
It should be pointed out that the large $\Delta=0.5$ eV enhances the Meissner effect, since $\delta \epsilon \propto 1/\Delta$.

To discuss the Meissner effect more precisely, the sample boundary should be taken into account. 
A further discussion of the role of the helical edge states\cite{KM1,KM2} in electron correlation is needed.


\begin{table}[htdp]
\caption{Parameters used for estimations. These are typical values for  Na$_2$IrO$_3$,\cite{Shitade} which is a honeycomb-layered topological band insulator with electron correlation.\\}
\begin{center}
\begin{tabular}{|c|c|c|c|c|}
\hline
$\Delta$ & $U$ & $d$ & $a$ & $v$
\\ 
\hline
0.5eV & 0.5eV & 10$\AA$ &  10$\AA$&$  3 \times 10^4$m/s 
\\
\hline
\end{tabular}
\end{center}
\label{table}
\end{table}%

The authors are grateful to N. Hatano, D. S. Hirashima, K.-I. Imura, S. Kurihara, S. Miyashita, 
T. Oka, Masahiro Sato, Masatoshi Sato, and A. Shitade for fruitful discussions and stimulating comments. 
J.G. is financially supported by a Grant-in-Aid for Scientific Research 
from Japan Society for the Promotion of Science under Grant 
No. 18540381.  J.G. is also supported by Core Research for Evolutional
Science and Technology (CREST) of Japan Science and Technology Agency.

\end{document}